\documentclass[pra,twocolumn]{revtex4-2}
  \usepackage{graphicx}
  \usepackage{amsmath}
  \usepackage{amssymb}
  \usepackage{makeidx}
  \usepackage{amsfonts}
  \usepackage{appendix}
  \usepackage{hyperref}
  \usepackage{mathrsfs}
  \usepackage{appendix}

  \hypersetup
  {
    colorlinks,%
    citecolor=black,%
    linkcolor=black,%
    urlcolor=black,%
  }

\makeindex

\begin{document}

\title {Converting between qubits of different forms}

\affiliation{Department of Physics $\&$ Astronomy, Seoul National University, Gwanak-ro 1, Gwanak-gu, Seoul 08826, Korea}
\author{Hyunseok Jeong}
\email{h.jeong37@gmail.com}
\affiliation{Department of Physics $\&$ Astronomy, Seoul National University, Gwanak-ro 1, Gwanak-gu, Seoul 08826, Korea}

\date{\today}

\begin{abstract}
{\bf A quantum bit encoding converter between qubits of different forms is experimentally demonstrated, paving the way to efficient networks for optical quantum computing and communication.}
\end{abstract}
\maketitle

\begin{figure*}[t]
    \centering
    \includegraphics[scale=0.6]{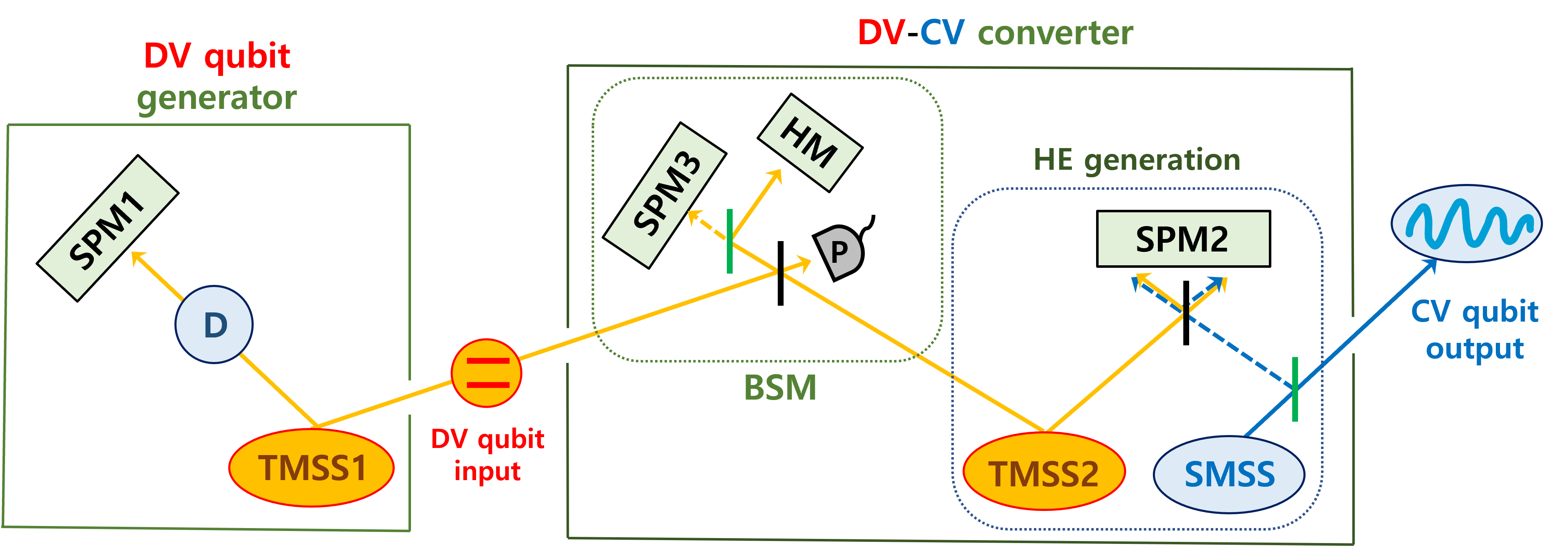}
    \caption{{\bf Schematic of a DV-CV converter~\cite{D2022}.} A DV qubit to be converted is generated using a two-mode squeezed state (TMSS1), a displacement operation (D) and a single-photon detector (SPM1). The DV-CV converter is composed of a hybrid entanglement (HE) generator and a Bell-state measurement (BSM). A hybrid entangled state is generated when the SPM2 clicks. If the results of the second single photon detector (SPM2) with a photodiode (P) and the homodyne detector (HD) indicate that a single photon is detected, the DV qubit is converted to a CV qubit. Black vertical bars indicate balanced beam splitters while green ones are unbalanced beam splitters.}
    \label{fig:mhd}
\end{figure*}

Quantum states of light are useful tools for quantum information technologies. They are particularly advantageous for quantum communication, networks, and metrology. They are also considered as a competitive candidate for building scalable architectures for quantum computation.

There are several options for how to create and encode qubits from light. A popular approach is the use of discrete variables (DVs), whereby polarization or photon-number degrees of freedom are often employed for qubit encoding. For example, the horizontal and vertical polarizations of a single photon (dual-rail encoding) \cite{M1989,KLM}
are typically used as well as the vacuum and single-photon excitation of a quantized light mode (single-rail encoding) \cite{LK2001,LR2002}.  

However, a limitation of DV encoding for photonic quantum information applications is that the cost for multiple-qubit operations to work deterministically is expensive. A representative example of this is the Bell-state measurement that plays a crucial role in optical quantum computation and communication \cite{L1999,C2001}. Hence, with DV encoding of light it is hard to efficiently correct errors during quantum computing and the efficiency of quantum repeaters that are critical for quantum communication protocols is degraded.

As an alternative, it is also possible to view the structure of a quantized field of light in a continuous-variable (CV) space \cite{BL05}. This opens diverse possibilities to encode, process, and measure quantum information. As a specific example, qubit encoding for quantum computing in terms of two coherent states of opposite phases $|\pm\alpha\rangle$ (and their superpositions such as $|\alpha\rangle\pm|-\alpha\rangle$ have been investigated as a way to overcome the limitations of DV encoding \cite{JK02,Ralph03}. Such qubits of superposed coherent states are s ``cat states'' as an analogy of Schrodinger’s cat paradox \cite{Sch,OJ2007}. They enable one to perform a nearly deterministic Bell-state measurement and two-qubit gate operations in a simpler way \cite{Jeong01,JKpuri2002}. On the other hand, their undefined photon numbers make them relatively more sensitive to detection inefficiency and decoherence caused by photon loss.

Now writing in Nature Photonics \cite{D2022}, Darras {\it et al.} report a qubit converter that can teleport a DV qubit to a cat-state qubit. Indeed, it is known that combining the DV and CV approaches would provide powerful tools for photonic quantum computing \cite{LJ2013,Omkar2020} and communication \cite{Sheng2013,Lim2016} as it enables one to take advantages of both the approaches. Earlier work has already experimental demonstrated DV-CV hybrid entanglement \cite{Jeong2014,Morin2014} and proof-of-principle demonstrations of CV to DV \cite{U17} and DV to CV \cite{S18} teleporters. Along these lines, the experiment of Darras {\it et al.} is another major step forward towards efficient networks for quantum computing and communication.

The first part of the authors’ experiment is to generate a DV qubit using a two-mode squeezed state (TMSS1), a small displacement operation (D), and a single-photon measurement (SPM1) as shown in Fig. 1. A two-mode squeezed vacuum contains the same number of photons in each mode. By performing a small displacement operation and a subsequent single-photon measurement on one of the two entangled modes, the other field mode is effectively projected to a superposition of the vacuum and single photon, i.e., a DV qubit in the single-rail encoding is created. By changing the strength of the displacement operation, the authors were able to generate a set of six DV qubits of different coefficients.

The second parallel part is to generate a hybrid entangled state, i.e., an entangled state between a DV qubit and a cat-state qubit. For this, another two-mode squeezed state (TMSS2) and a single-mode squeezed state (SMSS) are used together with the second single photon measurement unit (SPM2) as seen in Fig.~1.

Here, two points should be noted. First, only low-number-state terms are dominant in the TMSS2 because its degree of squeezing (and thus its average photon number) was small. Second, a single-mode squeezed state, as far as its squeezing degree is small, well approximates an even cat state in the form of $|\alpha\rangle+|-\alpha\rangle$ that contains only an even number of photons; this is the case for the SMSS in this experiment.

A small fraction of the SMSS and one part of the TMSS2 are mixed at a balanced beam splitter and the SPM2 is performed. When a single photon is successfully measured at the SPM2, it is impossible to know whether this single photon comes either from the TMSS2 or from the SMSS. This indistinguishability results in a superposition of two possible cases as follows. If the measured single photon came from the SMSS, this means that one photon was effectively subtracted from the SMSS. The SMSS would then be changed to a good approximation of an odd cat state, $|\alpha\rangle-|-\alpha\rangle$, that contains only an odd number of photons. Meanwhile, the vacuum part is the most dominant term for the TMSS2. Therefore, in this case, the remaining state would be approximately “a vacuum state and an odd cat” such as $|0\rangle(|\alpha\rangle-|-\alpha\rangle)$. In the other case, i.e., if the detected single photon came from the TMSS2, the SMSS remains the same while the single-photon detection of SPM2 affects the TMSS2 part that makes the remaining state to be “a single-photon state and an even cat” as $|1\rangle(|\alpha\rangle+|-\alpha\rangle)$. Thus, the total resulting state is in the form of $|0\rangle(|\alpha\rangle-|-\alpha\rangle)+|1\rangle(|\alpha\rangle+|-\alpha\rangle)$ [9] that is hybrid entanglement between the DV qubit and the cat qubit. In this way, hybrid entanglement is approximately generated using two squeezers, beam splitting, and a single-photon measurement.

The remaining critical part is to perform a Bell-state measurement that accomplishes the quantum teleportation protocol from the DV to cat-state qubits. It is highly demanding to identify all four Bell states \cite{L1999,C2001}. 
The Bell-state measurement for the teleportation from a single-rail DV qubit to a cat-state qubit can be performed using a 50:50 beam splitter and two single-photon detectors \cite{J16}. This identifies two of the Bell states.

The authors’ scheme efficiently detects only one of the Bell states by measuring a single photon using the SPM3 with a beam splitter and a photodiode (P); in order to confirm the result of ``only one photon" detected at the SPM3, they use a homodyne measurement (HM) with an additional unbalanced beam splitter as depicted in Fig. 1. At the unbalanced beam splitter, a small portion of signal is sent to the SPM3 for a single-photon measurement while the HM of the other side is to approximately confirm that the remaining part is the vacuum. A success is heralded both when the SPM3 clicks and at the same time when the outcome of the HM is ‘no photon’ within a small window. This limits the success probability of the protocol.

The authors performed quantum state tomography to reconstruct the input and output qubits and compared them to check how the teleportation operated. They obtained about 80\% of fidelity between the input and output qubits, which is over the classical bound obtained for their states ($\sim$74\%). They also carried out quantum process tomography to confirm that the process fidelity ($\sim$58\%) is over the known classical threshold 1/2.

Unlike a previously demonstrated DV to CV teleporter \cite{S18}, the authors’ experiment does not rely on post-selection of data. This makes the authors’ converter more practical for combining with external circuits and networks. On the other hand, approximate features of the entanglement generation and intrinsically low success probabilities are still limitations of the protocol. 

Since both the DV and CV approaches each have their own merits, an important direction for the future development of the optical quantum information technologies is to combine the two encodings in a same optical circuit or network \cite{Andersen15}. The development of convenient and practical techniques for converting between the different forms of qubits is therefore crucial. Thus, this work will stimulate further exciting research and development for combining diverse encoding methods with light fields.

\end{document}